\documentclass[eqnl]{aa}            % Option2 : Final A+A version (2 columns)
\input psfig.sty
\def\simless{\mathbin{\lower 1pt\hbox
   {$\spose{\raise 5pt\hbox{$\char'074$}}\char'430$}}}
\def\simgreat{\mathbin{\lower 1pt\hbox
   {$\spose{\raise 5pt\hbox{$\char'076$}}\char'430$}}}
\def\simgreat{\gapp}
\def\simless{\lapp}
\def\lapp{\mathbin{\raise2pt \hbox{$<$} \hskip-9pt \lower4pt \hbox{$\sim$}}}
\def\gapp{\mathbin{\raise2pt \hbox{$>$} \hskip-9pt \lower4pt \hbox{$\sim$}}}
\begin{document}

   \title{A multi-wavelength test of the FR~I - BL Lac unifying model
%   \protect\\
         }
   \titlerunning{A multi-wavelength test of the FR~I - BL Lac unifying model}

  \author{E. Trussoni
           \inst{1}
   \and  A. Capetti 
           \inst{1}
   \and  A. Celotti 
           \inst{2}
   \and  M. Chiaberge
           \inst{3}$^{,}$\inst{4}
   \and   L. Feretti
      \inst{4}
    }
   \offprints{E. Trussoni \\ (trussoni@to.astro.it)}

   \institute
         {Istituto Nazionale di Astrofisica (INAF) - Osservatorio Astronomico 
       di Torino, Strada Osservatorio 20, I-10025 Pino Torinese (TO), Italy
    \and S.I.S.S.A., Via Beirut 2--4, I-34014 Trieste, Italy
    \and Space Telescope Science Institute, 3700 S. Martin Drive, Baltimore,
MD 21218, USA
    \and  Istituto di Radioastronomia del C.N.R., Via  Gobetti 101, I-40129 
Bologna, Italy    
}
   \date{Received ... / accepted ...}

   \abstract{We collect multi-wavelength measurements of the nuclear emission
of 20 low luminosity FR~I radio-galaxies to test the viability of the FR~I - 
BL Lac  unifying model.  Although poorly sampled, the Spectral Energy 
Distributions (SED) of FR~Is are consistent with the double peaked shape 
characteristic of BL Lacs.  Furthermore while the distribution of the FR~Is 
in the broad-band spectral index planes shows essentially no overlap with the
regions where HBL and LBL are located, this can be simply due to the effects 
of  relativistic beaming.  
More quantitatively, deriving the beaming Doppler factor of a given
radio-galaxy from its X-ray luminosity ratio with respect to BL Lacs
with similar extended radio luminosity, we find that i) the luminosity
in all bands, ii) the value of the spectral indices, iii) the slope of
the X-ray spectrum, iv) the overall SED shape, may be all
simultaneously reproduced. However, the corresponding jet bulk Lorentz 
factors are significantly smaller than those derived for BL Lacs from other
observational and theoretical considerations. This suggests 
to consider a simple variant of the unification scheme that allows for the 
presence of a velocity structure in the jet.
}

\maketitle

\keywords{galaxies:active - jets - nuclei - BL Lacertae
objects:general
}

\section{Introduction}
The unifying model for low power radio--loud Active Galactic Nuclei
predicts that BL Lacs and Fanaroff-Riley I (FR~I, Fanaroff \& Riley
1974) radio galaxies constitute the same class of objects, the 
difference being the orientation of their relativistic jets with
respect to the line of sight.  This scenario has been mainly supported (and
tested) through the comparison of extended, presumably isotropic,
properties of the two populations (see Urry \& Padovani 1995 for a
review).

However it is also possible to verify and quantify the model
parameters by estimating (or setting tight limits on) the nuclear emission
in radio galaxies and directly compare it with the corresponding emission
in BL Lacs. Recently the increased instrumental spatial and spectral 
resolution has
in fact allowed the nuclear contribution in radio galaxies
to be separated from the rest of the emission; 
there are now reliable nuclear flux estimates
also in the X--ray and optical bands. Emission at optical and X-ray
 frequencies
appears to well correlate with the radio core one, as shown by
e.g. Canosa et al. (1999), Hardcastle \& Worrall (1999, hereinafter
HW99) for the X-rays and Chiaber\-ge et al. (1999,
hereafter CCC99) for the optical case (as deduced from HST
observations of FR~I 3CR galaxies). These correlations suggest the
same non-thermal origin of radiation in all three bands, thus implying
little (if any) obscuration along the line of sight in FR~Is.
Furthermore, Capetti \& Celotti (1999, hereafter CC99) have been able
to find for a subsample of (only) five FR~Is that the optical core
luminosity decreases by increasing the angle of the jet axis with
respect to the line of sight, in agreement with the unifying scheme.

Following the same line of work, combining the information for individual
sources in the different bands, Capetti et al. (2000; hereinafter
paper I) considered the spectral energy distributions of the same
small sample analyzed in CC99, in the attempt to compare the whole
individual SEDs of FR~Is with those of BL Lac objects.  In particular they
considered the broad-band spectral indices ($\alpha_{\rm ro}$,
$\alpha_{\rm rX}$ and $\alpha_{\rm oX}$), which provide a good description of
the SED, and also the X-ray spectral
information, which provides a valuable tool to discriminate between the
(steep) synchrotron and (flat) inverse Compton X-ray emission in BL
Lacs. A similar broad-band approach has been later adopted also by
Hardcastle \& Worrall (2000) and Bai \& Lee (2000). A general
agreement with the unification model has been found through this
comparison, with indications of the presence of two broad peaks, which
characterize blazar spectra, also found in the FR~I SEDs (with hints of the
presence of High and Low frequency peak sources, as in BL Lacs, Giommi
\& Padovani 1995). A very clear example of double peaked SED is
that of Centaurus A (Chiaberge et al. 2001) in which both peaks
can be accurately identified.  However, it became clear that the
simplest scenario involving a one velocity jet was not adequate to
consistently reproduce the SEDs of BL Lacs and FR~Is via beaming
effects. In particular the emission in radio galaxies exceeded what
was expected by de-beaming (at large angles) the typical BL Lac
luminosity. It was therefore suggested the presence of a slower jet
component (e.g. a layer of the jet) dominating the emission at larger
angles (Chiaberge et al. 2000).

Given the relevance of such issues for both the unifying model itself
and the physics of jets, we decided to further explore them by
considering a significantly larger sample of sources. For this purpose
we extend here the analysis of paper~I by constructing the SEDs for
FR~Is from the 3CR radio catalog for which multi-band observations are
available and by comparing their SEDs with those of BL Lacs.  The radio
and optical data are those reported in CCC99. A few available
infrared and UV observations not yet reported in the literature
have been extracted from the HST archive. Most of the X-ray data have
been taken from the recent literature, but again we also present
here the analysis of Rosat pointed observations of a few objects which
have not yet been presented elsewhere.  From this selection we
obtained a final sample of 16 radio galaxies, 20 including those examined 
in paper~I.
From the study of this relatively large sample of objects we are
able to deduce some further implications concerning the validity of the
unified model.

In the following Sect. 2 we list the sources considered in this work
and outline their main properties at different wavelengths, with
particular emphasis on the X-ray observations.  In Sect. 3 we build
up the SED of our objects and we discuss their features and
compare them with those of BL Lacs, while in Sect. 4 we test the 
BL Lac/FR~I unification by considering the different beaming conditions.
The main implications of our
results on the validity of the unified scheme are summarized in the
last Sect.  5.  Throughout this work we adopt the following values
for the cosmological parameters: $H_{\rm 0} = 75$ km s$^{-1}$
Mpc$^{-1}$ and $q_{\rm 0} = 0.5$. Spectral indices are defined by
$F_{\nu} \propto \nu^{-\alpha}$.

\section{The sample and the multi-wavelength data}
The sources have been extracted from  the 3CR for which X-ray data are
available  and core  emission  has  been detected  in  the HST  images
(CCC99).   In this  list we  have not  considered M~87,  that  will be
discussed separately,  and 3C 84, that  appears to be  a very peculiar
object (sometimes it is classified as a BL Lac) and highly variable at
all frequencies. Conversely we have kept in the sample 3C 346, that in
most publications is classified as  a narrow line, FR~II radio galaxy.
We remark also that 3C 348 (Hercules A) has properties similar to both
FR~I and FR~II  radio galaxies and its total  radio luminosity is more
typical of FR~IIs than FR~Is.  We have also considered in our sample 3C
189, which is part of the B2 sample of radio galaxies recently studied
using HST  data by Capetti et  al.  (2001), although  this source is
not included in the revised version of the 3C catalog (3CR).

In the following we outline the main properties 
of our targets, discussing separately the data at radio and IR/optical/UV
frequencies (Table 1) and those at X-ray energies (Table 2).

\noindent
{\it Radio observations}. For the radio core luminosities we referred
to the VLA observations at 5 GHz of Giovannini et al.  (1988) and
Morganti et al. (1993), reported in CCC99.  For some sources VLBI data
also are available, however we prefer to use the lower resolution VLA
observations as they have been obtained on scales similar to the
measurements at the other wavelengths.  The flux variability of the
radio core is generally negligible for our sources. Fluctuations of
$\approx 20 \%$ have been found in 3C 66B (Hardcastle et al. 1996),
whereas in 3C 338, the most strongly variable radio galaxy known of
our sample, the core flux may vary up to $\approx 60 - 70 \%$
(Giovannini et al. 1998).

\noindent
{\it Optical, infrared and UV observations}. Most of the HST observations in 
CCC99  were performed with the WFPC2 using the filter F702W; for
a few sources  data with different filters are also available.
In the public archive at STSCI observations with the HST/NICMOS are 
found at infrared frequencies for 3C 272.1 with the broad-band filters 
F110W, F160W and F205W, and for 3C 317 and 
3C 338 with  F160W. The analysis of these data have been performed as 
in paper I. UV images of 3C 66B and 3C 346 have been 
taken with the HST/FOC (at $\lambda = 4440$ \AA$\;$  and 3440 \AA) and 
HST/STIS (at $\lambda = 1400$ \AA), respectively.
All these observations have been performed in the years 1994 - 1996.
So far repeated HST observations are available only for a few sources. 
In particular, 3C 66B shows a decrease in flux by a factor $\approx$ 2 in 
the optical between 1995 and 2000, while 3C~317 has revealed strong 
variability in the UV (by a factor $\sim$ 10) between 1994 and 1999 
(Chiaberge et al. 2002). Other UV data of 3C~66B, 3C~78, 3C~317 
and 3C~338 are taken from Chiaberge et al. (2002).

All the fluxes have been corrected for the galactic extinction, but 
some amount of local absorption may be present. 
In Paper I we estimated that, based on the small dispersion of the 
radio/optical correlation for FR~I nuclei, such local extinction does not
exceed 2 magnitudes in the optical. 
This result is confirmed by recent UV observations 
(Chiaberge et al. 2002), which implies an underestimate of the luminosity of
by factor $\simless \, 6$ (this is probably the case in 3C 272.1 where a
central dust lane is present, CCC99). However, this is not critically 
relevant  for the construction of the SED (for details see paper I). 

\begin{table*}
\caption{The sources and their core luminosities in the radio,
$L_{\nu {\rm r}}$ ($\times 10^{23}$  W Hz$^{-1}$), and 
IR/optical/UV bands, $L_{\nu {\rm IR}}$, $L_{\nu {\rm o}}$ and $L_{\nu 
{\rm UV}}$
($\times 10^{19}$  W Hz$^{-1}$)}
\begin{tabular}{|c|c|c|c|c|c|c|c|}
\hline
Source   & Alt. name & z & Envir. & $L^{\rm a}_{\nu {\rm r}}$ &  $L_{\nu {\rm IR}}$ &
 $L^{\rm e}_{\nu {\rm o}}$ & $L_{\nu {\rm UV}}$  \\
\hline  
3C~~29     &          & 0.0438 &  A 119  &  3.5  &                & 4.1             &    $^*$                     \\
3C~~31     &  NGC 383 & 0.0169 & Arp 331 &  0.51 &                & 1.6             &                             \\
3C~66B     &          & 0.0215 &         &  1.6  &                & 8.8             & 3.5$^{\rm h}$,  2.5$^{\rm i}$, 1.35$^{\rm m}$  \\
3C~~78     & NGC 1218 & 0.0288 &         &  15   &                & 96, 85$^{\rm f}$&  1.44$^{\rm m}$           \\
3C~83.1    & NGC 1265 & 0.0251 & Perseus &  0.26 &                & 0.42            &                           \\
3C~~189    & NGC 2484 & 0.0413 &         &  6.5  &                & 25$^{\rm g}$    &             \\     
3C~272.1   &   M 84   & 0.0037 & Virgo   & 0.047 & 1.0$^{\rm b}$, 0.95$^{\rm c}$, 1.1$^{\rm d}$ & 0.37$^{\rm g}$, 0.13$^{\rm f}$ & \\
3C~277.3   &          & 0.0857 & Coma A  &  1.8  &                & 3.7,  5.4$^{\rm f}$   &                   \\
3C~288     &          & 0.246  &         &  39   &                & 15.1            &                          \\
3C~296     & NGC 5532 & 0.0237 &         &  0.84 &                & 0.65           &                    \\
3C~310     &          & 0.0535 &         &  4.5  &                & 3.6            &                   \\
3C~317     &          & 0.0342 & A 2052  &  8.9  &  27$^{\rm b}$  & 5.1$^{\rm g}$, 3.8$^{\rm f}$ & 0.22$^{\rm m}$        \\
3C~338     & NGC 6166 & 0.0303 & A 2199  &  1.9  & 2.6$^{\rm b}$  & 3.0            &  0.14$^{\rm m}$      \\
3C~346     &          & 0.162  &         & 120   &                & 241            & 29$^{\rm l}$   \\
3C~348     &          & 0.154  & Her A   &  4.9  &                & 8.1            &               \\
3C~442     & NGC 7237 & 0.0262 & Arp 169 & 0.027 &                & 0.23,  0.13$^{\rm f}$  &           \\    
3C~449     &          & 0.0171 &         &  0.21 &                & 2.5            &    $^*$          \\
\hline
\end{tabular} 
\vskip 0.1 true cm 
\noindent 
$^{\rm a}$ 5 GHz;
$^{\rm b}$ 16000 \AA;
$^{\rm c}$ 11000 \AA;
$^{\rm d}$ 20500 \AA;
$^{\rm e}$ 7020 \AA;
$^{\rm f}$ 5500 \AA;
$^{\rm g}$ 8140 \AA;
$^{\rm h}$ 4440 \AA;
$^{\rm i}$ 3400 \AA;
$^{\rm l}$ 1400 \AA; 
$^{\rm m}$ 2500 \AA;
$^*$ we did not use the HST data in the UV band for 3C~29 and 3C~449 
since the nuclear emission of these sources is clearly affected by 
obscuration from extended (kpc--scale) dusty structures 
(Chiaberge et al. 2002).
\end{table*}

\noindent
{\it X-ray observations}. We discuss in more detail the
observations in the X-ray band that are particularly crucial for us to be able
to define the structure of the SED. 

We must point out that in instruments with poor imaging capability the nuclear
X-ray flux may be contaminated by spurious sources, like X-ray binaries, 
jets, etc. However the  correlation between the radio and X-ray core
luminosities (Canosa et al. 1999, HW99), suggests that in general the jet 
emission should provide the most relevant contribution to the nuclear flux.
We remind in particular that for the radio galaxies of our sample the 
integrated contribution from binaries  to the total emission is not relevant 
(see e.g.  Fabbiano 1989). Furthermore from Chandra data it 
has been deduced that emission from X-ray jets is not the dominant source of 
high energy photons in these radio galaxies (Hardcastle et al. 2001, Worral 
et al. 2001). It has also  been proposed that advection dominated disks (ADAF) 
could contribute to  the core emission from radio to hard X-ray energies 
(see e.g. Narayan \& Yi 1995, Di Matteo et al. 2000, 2001).  In normal
elliptical galaxies it has been  excluded a relevant contribution from the 
ADAF to the nuclear luminosity (Loewenstein et al. 2001).  Chandra 
observations suggest that this should be also the case for radio galaxies
(Di Matteo et al. 2002, Pellegrini et al. 2003). The above considerations
leave us with extended hot galactic gas as the main source of contamination.

The data on our sources are quite
heterogeneous, as they have been collected from different instruments
and satellites. In all cases the luminosities have been re-scaled to the
soft X-ray energy range of Rosat (0.1 - 2.4 keV). Most of the data
reported here have been extracted from recent publications, while we
directly analyzed the archival Rosat data of 3C 29, 3C 66B, 3C 189, 3C
277.3 and 3C 317 not yet discussed in the literature.

Very recent Chandra data have been collected for 3C 31{\footnote 
{We reconsider here this source that was previously analyzed in paper I
using Rosat data}} (Hardcastle et al. 2002), 3C 66B (Hardcastle
et al. 2001), 3C 189 (Worral et al. 2001), 3C 272.1 (Finoguenov \& Jones
2001), 3C 317 (Blanton et al. 2003) and 3C 338 (Di Matteo et al. 2000). 
In all sources a point-like
core has been detected, with  flux almost unabsorbed locally 
($N_{\rm H}  \simless \,$ few units of $10^{21}$ cm$^{-2}$), however only for 
3C 31, 3C 66B and 3C 272.1 is there a good evaluation of the spectral slope.

For six objects the core emission has been searched for 
through spectral fits with two--component models: thermal (from the
environment) + power law (from the nucleus). A positive detection has
been obtained for 3C 78 (BeppoSAX, Trussoni et al. 1999), 3C 346 (ASCA
and Rosat, Worral \& Birkinshaw 2000) and 3C 348 (BeppoSAX, Rosat and
ASCA, Siebert et al. 1999, Trussoni et al. 2001), although
with poor,
if any, estimate of the spectral index.  From our spectral analysis
of the Rosat/PSPC observations of 3C 29, 3C 277.3 and 3C 317 only an upper
limit to the core emission has been deduced for the first two radio
galaxies while for 3C 317 non-thermal flux has been detected with very
flat spectrum ($\alpha_{\rm X} \approx 0.1$).  For all those sources with no
estimate of the spectral index, the X-ray luminosities have been
deduced assuming $\alpha_{\rm X}=1$.

Finally, for the remaining six radio galaxies (3C 83.1, 3C 288, 3C 296, 3C 310 
and 3C 449) we referred 
to the results of HW99 who, exploiting the imaging capabilities of the HRI 
and PSPC of Rosat, looked for the  presence of point-like components coinciding
 with the active nuclei. As a small thermal region may be present in the 
central regions of radio galaxies the nuclear flux  can be overestimated, 
mainly  for distant sources (as e.g. in 3C 465, at the center of cluster
 A 2634, see  paper I). However 
the correlation between the radio and X-ray core luminosities suggests 
that in most of the detected cores the non-thermal component cannot be 
a negligible fraction of the total emission. In any case the values of 
the luminosities  will be considered as upper limits of their nuclear emission 
(marked with a star in Table 2). 

For $\approx 2/3$ of the targets repeated X--ray observations are
available.  However in  most cases the count rate is  too low (or only
upper limits exist) to obtain  information on the possible presence of
variability.  From ASCA  and  Rosat  data a  slight  decrease of  the
emission has been found in  3C 346 over $\approx$ 1.5 yr. In contrast,
from two PSPC and one Chandra observation it turns out that the X-ray
core  flux from 3C  66B has  decreased by  a factor  $\approx 4$  in 9
years. A similar trend has been found for 3C 317,  although the large
statistical errors do not allow a definite conclusion. Out of the repeated 
observations  of these three targets we use those closer in time to 
the HST pointings.

More details on the observations and the X-ray 
properties of the other objects not discussed in HW99 
are reported in the Appendix.

\begin{table}
\caption{ Data on X-ray nuclear emission used for the SED 
(see the Appendix for details)}
%%%\begin{tabular}{|l c c c c |}
\begin{tabular}{|c|c|c|c|}
\hline
Source    & $L^{\rm a}_{\nu {\rm X}}$ & $\alpha_{\rm X}$ &  Instrum.(refer.)   \\
\hline
3C~~29    & $5.5^{\star}$       & 1.0                   & PSPC (1)      \\
3C~~31    & 0.46                & $0.5^{+0.3}_{-0.2}$     & Chandra (2)  \\
3C~66B    & $9.5^{+0.2}_{-0.9}$ & $0.3^{+0.3}_{-0.4}$   & PSPC (1) \\
3C~~78    & $32^{+10}_{-17}$    & 1.0                   & SAX (3)      \\
3C~83.1   & $3.6^{\star}$       & 1.0                   & PSPC (4)       \\
3C~~189   & 14                  & $1.1^{+0.1}_{-0.4}$   & Chandra (5)  \\
3C~272.1  & $0.051^{+0.003}_{-0.003}$ & $1.3^{+0.1}_{-0.1}$   & Chandra (6)       \\
3C~277.3  & $<5.3$              & 1.0                   & PSPC (1)        \\
3C~288    & $<2100$             & 1.0                   & PSPC (4)        \\
3C~296    & $6.0^{\star}$       & 1.0                   & HRI (4)        \\
3C~310    & $16^{\star}$        & 1.0                   & HRI (4)        \\
3C~317    & $30^{+21}_{-23}$    & $0.1^{+0.3}_{-0.7}$   & PSPC (1)        \\
3C~338    & $0.53^{+0.11}_{-0.11}$ & $0.5^{+0.5}_{-0.4}$   & Chandra (7)        \\
3C~346    & $618^{+110}_{-160}$    & $0.7^{+0.2}_{-0.2}$   & ASCA (8)      \\
3C~348    & 234                 & 1.0                   & PSPC, ASCA, SAX (9)  \\
3C~442    & $<1.9$              & 1.0                   & HRI (4)       \\
3C~449    & $0.89^{\star}$      & 1.0                   & PSPC (4)        \\
\hline
\end{tabular} 
\vskip 0.1 true cm 
\noindent
$^{\rm a}$ $\times 10^{16}$ W Hz$^{-1}$ (at 1 keV). For each object $N_{\rm H}$
is fixed to its galactic value, except for 3C 272.1 where it is deduced from the fit 
($N_{\rm H} = 2.7^{+0.3}_{-0.3} \times 10^{21}$ cm$^{-2}$). In the 
sources marked with $^{\star}$ contamination from thermal emission may 
be present
\vskip 0.1 true cm
\noindent
References: (1) this work; (2) Hardcastle et al. (2002); (3) Trussoni et al. 
(1999); (4) HW99; (5) Worral et al. (2001); (6) Finoguenov \& Jones (2001);  
(7) Di Matteo et al. (2000);  (8) Worrall \& Birkinshaw (2001); 
(9) Siebert et al. (1999), Trussoni et al. (2001).
\end{table}

\section{Results}
As already mentioned in the Introduction, the correlations found
between the luminosities of the nuclei of FR~Is in different spectral
bands, from radio to X--ray energies, suggest a common non-thermal
origin for the emission form these nuclei, occurring in the inner
parts of the relativistic jets.  Our aim is to use the data 
to compare the multi-wavelength properties of the nuclear emission
of FR~Is with those of BL Lacs, in order to  derive information
on the jet properties. In particular we examine 
the Spectral Energy Distribution, the intensity of the X-ray emission and
the values of the broad-band spectral indices. 

\subsection{The Spectral Energy Distribution}

The main characteristic of BL Lac SEDs is that they can be well
described as composed by two broad peaks. The lower energy one, most
likely due to synchrotron emission, is located in the mm-IR or 
UV-soft/X-ray bands in Low and High peak BL Lacs, 
respectively (HBL and LBL, in the
terminology introduced by Giommi \& Padovani 1995). The higher energy
component comprises the X-ray and $\gamma$-ray range and is possibly
due to inverse Compton emission.  Therefore, it is important
to look for the presence of such two components also in the SEDs of FR~Is, 
to investigate if
indeed their nuclear emission is dominated by non-thermal radiation,
only de-beamed with respect to BL Lacs.  Positive indications of the
compatibility of the behaviour of such a spectral shape with the (although poorly
sampled) SEDs of FR~Is have been found in paper~I. Here we further investigate
the SED shape for a larger number of sources, and with better sampled energy
distributions, and try to locate the positions of the peak of the
synchrotron component.

The SEDs of our  objects are shown in Fig. 1.  All  of them present an
increase  of  the  emission   from  radio  up  to  $\approx$  infrared
wavelengths,  while significant  differences can  be seen  at higher
frequencies. Indications of the shape of the SEDs and on the location of the
emission  peak can  be inferred  from  the relative  intensity of  the
optical  and X--ray emission  and -  as already  mentioned -  from the
X--ray spectral index.  Unfortunately, for half of the sources we have
only an  upper limit to  the nuclear non-thermal X--ray  emission (see
Table 2).  We do not further  consider here the sources  for which the
X--ray limit is  not significant (3C 83.1, 3C 288, 3C  296, 3C 310 and
3C 442), whereas in three cases (3C 29, 3C 277.3 and 3C 449) the limit
is still useful to provide constraints on the SED.

In 3C 78, 3C 272.1 and 3C 449, the X-ray luminosity is significantly 
lower than that of the optical component. This suggests
the existence of a minimum approximately
around  X-ray energies, in the region of transition from
synchrotron to Compton emission and it is supported by the steep
X-ray spectrum of 3C 272.1. 

In five radio galaxies (3C 31, 3C 66B, 3C 317, 3C 338 and 3C 346) 
the X-ray spectrum is rather flat (the SED is rising
at these energies) suggesting that the synchrotron peak must 
be located at lower frequencies. Even though the overall shape of the 
SED may be affected by spurious emission in the IR/optical band 
(e.g. by dust emission) the above interpretation seems to be confirmed 
by the comparison of the SED slope in the  IR/optical/UV spectral 
region. In fact in all these cases the slope is decreasing, 
suggesting for the presence of a minimum and for a location of
the synchrotron peak at $\sim $ IR frequencies, or lower.  The SEDs 
of these sources are then qualitatively similar to the SEDs of LBL:
the X-ray emission is plausibly of Compton origin with a flat spectrum.
Very similar properties are present in the SED of 3C 270 (see paper
I). 

Conversely, the data for 3C 29, 3C 189, 3C 277.3 and 3C 348 
(similar to 3C 465 discussed in paper I) do not allow to
characterize the shape of their SEDs.

While these findings generally support the presence
of a double peaked SED of FR~Is, it is necessary to stress
that the luminosity scatter of these objects ranges over $\sim$ 2-3
order of magnitudes in the radio and X-ray bands, 
and it is impossible to derive any trend with the source power.

\subsection{Comparison between the X-ray emission in FR~Is and BL Lacs}

Measurements of the nuclear emission  in FR~I radio galaxies and their
comparison with  the  luminosities  of BL Lacs provide  us with a
direct measure  of the amount  of beaming affecting the  two different
classes  of  objects. For this  test we must consider  
objects of  similar large scale properties,   
which do  not  depend  on   orientation.  In
particular, they must share   a  common  range  of  extended  radio
luminosity as this is linked to the energy carried by the relativistic
jet, which is plausibly at the origin of the nuclear emission.

The same method  has been applied both  in the optical and  radio bands to
the 3C  and the B2  samples of FR~I  radio galaxies (Chiaberge  et al.
2000, Capetti et  al. 2001). The luminosity ratio  between FR~Is and BL
Lacs  was found to be $L_{\rm BL}/L_{\rm FR~I}   \approx  10^{2.5}$  and  
$L_{\rm BL}/L_{\rm FR~I} \approx 10^{3.9}$  in the radio and optical  bands, 
respectively.  These authors deduced  that the difference in nuclear 
luminosity (in both bands) between  BL Lacs and FR~Is can  be accounted 
for by  a single amplification  factor, taking into account the scatter 
of the data, with corresponding jet bulk Lorentz factors 
$\Gamma \approx 4 - 6$ (Chiaberge et al. 2000).  Similar values in FR I objects
have been deduced by Giovannini et al. (2001) from radio data. However
these Lorentz factors are significantly smaller than those derived for 
BL Lacs from other observational and theoretical considerations. Namely, 
the variability and time lags at different frequencies, the spectral 
fittings of  the overall SED,  and the detection of $\gamma$ -ray emission 
constrain $\Gamma \simgreat 15$ (Dondi \& Ghisellini 1995, Ghisellini 
et al. 1998, Tavecchio et al. 1998).
 
Chiaberge et al. (2000) have shown that the presence of a velocity
structure in the jet (a fast spine surrounded by a slower layer) could
overcome this difficulty. In such a picture, observers at different viewing
angles see different components of the jet: the fast
spine  dominates  the  emission  in  BL Lacs,  whereas  the  slower  layer
dominates the emission observed in  FR~Is.  We remark  that such a model
has been suggested, from radio observation,  for 3C 31 (even though on a
much larger scale; Laing \& Bridle 2002). 

Here we  adopt the same  strategy for the X-ray  nuclear luminosities.
In Fig. 2 we show our samples  of FR~Is and BL Lacs in the plane formed
by  $L_{\nu {\rm X}}$  against  $L_{\nu {\rm r},  {\rm  ext}}$ (i.e.   the
nuclear  X-ray luminosity  vs  the extended  radio  luminosity at  1.4
GHz). Note  that the two  classes of BL  Lacs  share the same  range of
X--ray  luminosity independently  on  $L_{\nu {\rm r},  {\rm  ext}}$ 
($L_{\nu {\rm X}}  \sim
10^{27}$ erg s$^{-1}$ Hz$^{-1}$, with  a scatter of $\sim \pm$ 1 order
of magnitude).

The  presence  of several  upper  limits  in the FR~Is
makes  the estimate  of  the
relative beaming  factor in the X-ray  band less accurate  than in the
radio and optical  bands.
Although a correlation between $L_{\nu {\rm X}}$  and
$L_{\nu {\rm r},  {\rm  ext}}$  in the FR~Is cannot be firmly established,
we followed the method used in Capetti et al. (2001) 
and Chiaberge et al. (2000), and  fitted a regression line  
to the data, using the survival analysis package {\it ASURV}
(Isobe  \& Feigelson  1985,  1986).  In this way, we derived
that at $L_{\nu  {\rm r},  {\rm ext}}  \sim
10^{32}$  erg s$^{-1}$  Hz$^{-1}$, which  is the  median value  for our
sample,  the  corresponding  X--ray  luminosity  is  $L_{\nu  {\rm X}}  \sim
10^{23.5}$ erg  s$^{-1}$ Hz$^{-1}$. Thus the  luminosity ratio
between FR Is and BL  Lacs is of  $\sim$ 3.5 orders of  magnitude. This
corresponds to a ratio between the Doppler factors of BL Lacs and FR~Is
$R_{\delta} \equiv \delta_{\rm BL}/\delta_{\rm FR~I}=  (L^{\rm BL}_{\nu
{\rm X}}/L^{\rm FR~I}_{\nu   {\rm X}})^{1/(p+\alpha)}$   between   10   and   25   for
$\alpha_{\rm X}=1.5-0.5$,  respectively,  and $p=2$  (as  appropriate for  a
continuous  jet){\footnote {This relation between the luminosities and 
the Doppler factors strictly holds for synchrotron and Self-Synchro-Compton 
(SSC) emissions, while for External Compton emission (EC) we should shift 
$1/(p+\alpha)  \rightarrow 1/(p+1+2 \alpha)$ (Georganopoulos et al. 2001). In 
such regime of  Compton emission it is  $0.5 \geq \alpha_{\rm X} \geq 1$, 
and for ratios of luminosities  of $10^{3.5}$  we overestimate $R_{\delta}$ 
by a factor $\sim 3.0$ if the emission mechanism is EC. This
in principle affects the beaming tracks of flat X-ray spectrum  sources
but does not critically modify our conclusions considering   
the scatter of the data in Fig. 2 ($\sim$ one order of magnitude).}}   
In  the  case of  a  relativistically moving  sphere
($p=3$) is $R_\delta \sim 6-10$, for the same range of $\alpha_{\rm X}$.  These
values are therefore compatible with  those found for both  the radio
and   the   optical   bands   ($R_{\delta}\sim   18-20$   for   $p=2$;
$R_{\delta}\sim 7-9$  for $p=3$), although  they have a  larger allowed
range, due to the uncertainty on the local X--ray spectral index.

If there is a velocity gradient across the jet, 
$R_\delta$ corresponds to an  ``effective'' beaming factor,
which accounts for the effects of the different jet speed components.
Therefore, the comparison of the nuclear luminosity of BL Lacs and FR~Is 
does not lead to a ``real''  value of a unique Lorentz factor of  the jet.
As  in the  case of the  radio and  optical bands, the inferred 
values of the bulk Lorentz factors are lower than those
obtained  for the BL Lacs from other arguments.
Indeed, by adopting the reference
value of  $\delta_{\rm BL} \approx \Gamma=15 \div 20$ (see Chiaberge et al. 2000, 
and references therein) and a viewing angle
for BL Lacs and FR~Is of $\theta = 1/\Gamma$ and $\theta
\sim 60^{\circ}$ respectively, we obtain $R_\delta$   $\sim 100 \div 200$.  
This value  of $R_\delta$ would translate
in a  ratio $L^{\rm BL}_{\nu {\rm X}}/L^{\rm FR~I}_{\nu {\rm X}} = 10^5-10^9$ for
the range  of parameters used  above ($p=2,3$; $\alpha_{\rm X}  = 0.5,1.5$).
Therefore, we outline  that also the X-ray data 
 are consistent with the possible presence of a velocity  structure in the jet.

A second deduction that can be made from the $L_{\nu {\rm r}, {\rm ext}}/
L_{\nu {\rm X}}$ plane is an estimate of $R_{\delta}$ for each individual
source comparing its X-ray luminosity with BL Lacs of similar extended radio luminosity.
However this comparison is meaningful only if the difference in X--ray luminosity at a
given extended luminosity can be ascribed mainly to a different
orientation of the radio galaxy. As
already noted in the optical and radio bands, it also the case
in the X-rays that we
have a clear indication that objects seen at a small angle with
respect to the line of sight are over-luminous.  In particular, all
sources with a resolved optical jet, an indication of a close jet alignment
(see Sparks et al. 1995, 2000), namely 3C 66B, 3C 78, 3C 189, 3C 264
and 3C 346, are among the brightest objects. At the same time we note
the low luminosity of 3C 338 ($L_{\rm BL}/L_{\rm FR~I} \approx 10^{4.5 \pm 1}$)
which suggests a largely de-beamed nuclear emission, consistent with a
jet basically projected on the sky plane, as deduced from radio data
($\theta \approx 85^{\circ}$, Giovannini et al. 2001). 
The spread in X-ray luminosity then appears to be dominated by orientation effects
rather than by intrinsic differences from source to source and therefore this
information can be used to constraint the ``beaming
tracks'' that will be discussed in the following section.

Finally,  
we note  that  on average  HBLs  are
characterized  by  lower extended  radio  luminosities  than LBL:  the
former  sample   extends  from  $\sim  10^{29}$   to  $10^{31.5}$  erg
s$^{-1}$Hz$^{-1}$,    the    latter    from   $\sim    10^{30}$    erg
s$^{-1}$Hz$^{-1}$ to  $10^{33}$ erg s$^{-1}$Hz$^{-1}$ (Chiaberge et al. 2000, 
and references therein).
As the SED's of these  two BL Lac classes are
different (Fossati et al. 1998),  it would be interesting to test if 
such a distinction also reflects to the FR Is. We might
also expect that radio-galaxies of low extended luminosity will show a
SED typical  of HBL,  while the highest  luminosity objects  should be
LBL-like.  Unfortunately, our FR~I  sample  is mostly  concentrated on  the
transition  region   across  LBL  and  HBL. The few  objects  with extreme 
luminosities and well defined SEDs
are 3C  272.1, with  $L_{\nu {\rm r},  {\rm ext}}$ typical  of HBL,  and the
powerful 3C 346 and 3C 348, with the same extended luminosity as LBL.
Although clearly not conclusive, these results do not contradict 
the trend observed in BL Lacs.

\subsection{Distribution of $\alpha_{\rm ro}$, $\alpha_{\rm rX}$ and $\alpha_{\rm oX}$}

Further information on the unification model can be obtained by considering 
the broad-band spectral indices $\alpha_{\rm ro}$, $\alpha_{\rm rX}$ and 
$\alpha_{\rm oX}$ of FR~I radio galaxies and comparing with those of BL Lacs
(Fig.~3; the  3CR sources of paper I are also included). 

It is evident 
that there is essentially no overlap of the distribution of the spectral
indices of FR~Is with the regions of these planes where HBL and LBL
are located. In particular,
with respect to HBL a large radio excess is present in radio galaxies 
(both $\alpha^{\rm FR~I}_{\rm ro}$ and $\alpha^{\rm FR~I}_{\rm rX}$ are greater
than $\alpha^{\rm HBL}_{\rm ro}$ and $\alpha^{\rm HBL}_{\rm rX}$ respectively).
Overall the values of $\alpha_{\rm rX}$ are
similar to those of LBL, but only few sources fall in the LBL
region as in most of the FR~I objects we have $\alpha_{\rm oX} <  1.1$
and $\alpha_{\rm ro} >  0.7$,
i.e. they show an excess of X-ray and radio emission 
with respect to LBL.

To establish whether the distributions shown in Fig.~3
are indicative of an intrinsic difference between the two classes of
objects it is essential to evaluate the effect of the relativistic
beaming. This point will be discussed in the following section.

\section{Testing the FR~I - BL Lac unified model}

Based on the results outlined in the previous section, it is now
possible to perform a more quantitative comparison between FR~Is and BL
Lacs. In particular, we can test if the information on the SED, spectral
indices and difference in luminosities of the two classes can be
simultaneously accounted for within the simplest unification scheme,
only allowing for the presence of a velocity structure in the jet that
is suggested by the observations (see Chiaberge et al. 2000, Capetti et
al. 2001).

We reiterate that from the luminosity ratio between a FR~Is and BL Lacs in 
a given band (see Sect. 3.2) we can estimate the ratio of the beaming 
factors $R_{\delta}$. At the same time, moving from a
large viewing angle to a jet more closely aligned with the line of
sight the SED is Doppler shifted to higher frequencies by a factor
$\propto R_{\delta}$. As the SED is not represented by a single power
law, the broad-band spectral indices transform as (Chiaberge et
al. 2000):
$$
\alpha^{\rm BL}_{ij} = \alpha^{\rm FR~I}_{ij} + (\alpha_i - \alpha_j)
{{{\rm log} R_{\delta}} \over {{\rm log} (\nu_j/\nu_i)}}
\eqno(1)
$$  

\noindent
where $i,j \equiv {\rm r,o,X}$. The above relation provides a `beaming
track', i.e. the path of a representative radio galaxy in the
$\alpha_{\rm ro}$ and $\alpha_{\rm rX}$ vs $\alpha_{\rm oX}$ planes as a function of
$R_{\delta}$. 
 
It is clear that a crucial ingredient to perform this
comparison effectively is the X-ray spectrum of the radio-galaxies.  This
information is available for eight of our radio galaxies (six from
Table 2 and 3C 264 and 3C 270 from paper I).  For the radio emission
we have assumed a spectral index $\alpha_{\rm r} = 0$.  For the optical band
only for few sources in the sample the photometry is sufficiently
accurate to allow a constrained enough estimate of $\alpha_{\rm o}$
($\approx 0.7 - 1.3$, see CC99). To be conservative we have assumed
the spectral index free to vary in the interval $0.5 \leq \alpha_{\rm o}
\leq 1.5$ (from their standard SED we expect $ \alpha_{\rm o} > 1$ and $
\alpha_{\rm o} \leq 1$ for LBL and HBL, respectively).
   
We then proceed to analyze the tracks for those sources with the best  
data, by estimating $R_{\delta_{\rm X}}$ from the ratio between their measured  
X-ray luminosity and the average luminosity of BL Lac objects (see Fig. 2)
and taking into account the X-ray spectral index. Referring in
particular to 3C 272.1, which has the best sampled SED, we find
$L_{\rm BL}/L_{\rm FR~I} = 10^{5\pm1}$ and $\alpha_{\rm X}=1.3$, which yields
$R_{\delta_{\rm X}} \approx 10^{1.5 \pm 0.3}$.  With this allowed range of values
for the ratio of the beaming factors estimated in the X-rays, it is
possible to check if, by adopting the same range for the optical and radio
band (i.e. $R_{\delta} = R_{\delta_{\rm X}}$), the resulting relative luminosities
are also consistent with a given class of BL Lac objects. 
The tracks of 3C 272.1 fall in the region of overlapping between HBL and
LBL (see the shadowed regions in Fig. 4, right panels).
Thus the `beamed' 3C 272.1 falls, in both planes, in the HBL
zone almost independently of the slope of the optical spectrum
(Fig. 4,
left panels). We have already seen that the X-ray spectrum of 3C 272.1
is rather steep and its SED has a minimum around the X-ray band, both
features typical of HBL. Finally, its extended radio emission is also
typical of HBL sources.  In summary there is a complete consistency
between all spectral and luminosity indicators, and any optical
extinction does not modify these conclusions.

The tracks of 3C 66B ($R_{\delta} \approx 10^{1.4 \pm 0.4}$,
$\alpha_{\rm X}=0.3$; Fig. 5)
`join' the LBL region, as expected by its flat X-ray spectrum and by the
shape of its SED, again independently of the optical spectrum. A similar
trend holds for 3C 346 for which a small amount of beaming is required
due to its high luminosity ($R_{\delta} \approx 10^{0.4 \pm 0.4}$;
Fig. 6).
The X-ray spectral slope ($\alpha_{\rm X}=0.7$) 
and the extended radio luminosity are typical of LBL. Indeed 
3C 346 appears as  a LBL-like radio galaxy in all diagnostic planes, 
provided the optical spectrum is quite steep 
($\alpha_{\rm o} \approx 1.5$), consistently with the standard SED of LBL. 

Referring to the sources studied in paper I, the analysis of the
beaming tracks confirm that 3C 270 ($R_{\delta} \approx 10^{1.5 \pm
0.4}$, $\alpha_{\rm X} = 0.7$), can be associated to a LBL, as we already
argued based on its SED.  As in the case of 3C 346, 3C 270 falls into
the LBL region only if its optical spectrum is quite steep ($\alpha_{\rm o}
\approx 1.5$) as indeed observed for this source.  For 3C 264
($R_{\delta} \approx 10^{0.7 \pm 0.3}$) we have an accurate 
measurement of
$\alpha_{\rm X}$ (= 1.45), while from the IR and optical data we can
estimate $\alpha_{\rm o} \approx 0.8 - 1$.  The diagnostic plots (Fig. 7)
confirm that this radio galaxy can be associated with a HBL.

More ambiguous is the case of 3C 189 ($R_{\delta} \approx 10^{1 \pm
0.3}$, not shown here) as neither its X-ray spectral slope ($\alpha_{\rm X} =
1.1^{+0.1}_{-0.4}$) nor its SED clearly identify it with a given class
of BL Lacs.  The beaming tracks of the spectral indices seem to favour
an association with a LBL, but the available data are not conclusive.

The quality of spectral data on the remaining sources of our sample 
do not allow us to explore their beaming tracks.

We must remind finally that in Eq. (1) the spectral slopes in the 
different bands are kept constant, which is in principle inappropriate 
taking into account that  $R_{\delta} \approx 10 \div 30$. This 
assumption may be critical only in the region of the optical peak and  
of the transition between the  Compton and synchrotron emission. In such 
a case the values of $\alpha_{\rm ro}$ and $\alpha_{\rm rX}$ could be misestimated 
by a quantity $\Delta \alpha_{ij} \simless 0.1$. However we have seen above 
that a similar amount of uncertainties affect the evaluations of $R_{\delta}$ 
from Fig. 2.

\section{Summary and Conclusions}

We combined measurements in the X-ray band with optical nuclear
luminosities derived from HST images and radio data taken from the
literature, to characterize the multi-wavelength properties of 20 FR~I
radio galaxies from the 3CR sample.  In the framework of the
unifying model for FR~I and BL Lac the results of the present analysis
can be summarized as follows:

\noindent
{\it 1.} The SED of the sources are not monotonic showing, as in BL Lacs,
peaks and minima of emission; in more than half of the objects this 
minimum occurs around the soft X-ray band.

\noindent
{\it 2.} A comparison of the X-ray luminosities of FR~Is and BL Lacs
indicates that a single amplification factor accounts for their
difference in luminosity in the radio, optical and X-ray band.

\noindent
{\it 3.} The dominant role  
of Doppler boosting is confirmed, while obscuration of the nucleus 
does not appear to be relevant in the FR~I/BL Lac unification.

\noindent
{\it 4.}  The distribution of FR~Is in the broad-band spectral index
planes shows essentially no overlap with the region typical of HBL and
LBL. However, we have shown that this behaviour is only apparent and
related to the effects of relativistic beaming and the BL Lac SED.

\noindent
{\it 5.} Based on these results we have performed, for the sources
with the best data, a more complete and quantitative comparison between
FR~Is and BL Lacs to verify the consistency between all spectral and
luminosity indicators. Namely we derived the beaming factor for 6 FR~I
sources (including also two objects analyzed in paper I) based on
their X-ray luminosity as compared to that of a BL Lac of similar
extended radio luminosity. From the estimate of this single free
parameter we predicted optical and radio luminosities and broad-band
spectral indices for its counterpart observed at a small angle from the
line of sight.  This provided us with an association of each FR~I
source with a given type of BL Lac, which is found to be in agreement
with the slope of the X-ray spectrum, the overall SED shape and the
extended radio luminosity of the corresponding BL Lac class.

Thus the differences between FR~Is and BL Lacs can be all accounted for
within the simplest unification scheme, where the nuclear FR~I emission
is non--thermal radiation from a relativistic jet, only allowing for 
the presence of a velocity structure in the jet. This is required 
for Centaurus A, for which very detailed
multi-band data up to GeV energies are available and the orientation
of the jet is known (Chiaberge et al. 2001). 
On the other hand the fact that a similar amount
of beaming occurs from radio to X-ray energies puts some strong
constraints on the jet dynamics. As it is likely that the X-ray and
optical emission regions occur closer to the central engine than the radio
emission by $\approx 2$ orders of
magnitudes, the jet structure would have to be preserved on these
length scales.

In conclusion let us briefly consider 
how the validity of this procedure to test the unified model could
improve with better observational information.  Apart the
availability of a larger sample of objects, multi-frequency
observations would be necessary to estimate the spectral slope at
IR--optical--UV wavelengths. Furthermore with respect to previous
X-ray data, Chandra is greatly improving our ability to find point-like
nuclei. However, in general, they are so weak that only in few cases we
had a satisfactory estimate of $\alpha_{\rm X}$ and $N_{\rm H,loc}$ (for this
purpose also data from XMM could be useful). Finally we point out that the
variability of the core fluxes at all frequencies may affect the
shape of the SED and the evaluation of the indices $\alpha_{ij}$ (this
may be quite crucial at X-ray energies, see e.g. the case of 3C
66B). A useful step forward to test the validity of the unified
scheme would be simultaneous multi-frequency observations.

\begin{acknowledgements}
This work has been partially supported by the Italian Space 
Agency (ASI) and the Italian Ministery for Education and Research (MIUR, 
cofin 2001/028773). The authors wish to thank M. Georganopoulos for his 
useful comments and suggestions.
\end{acknowledgements}

%Figure 1
\begin{figure*}
\centerline{\psfig{figure={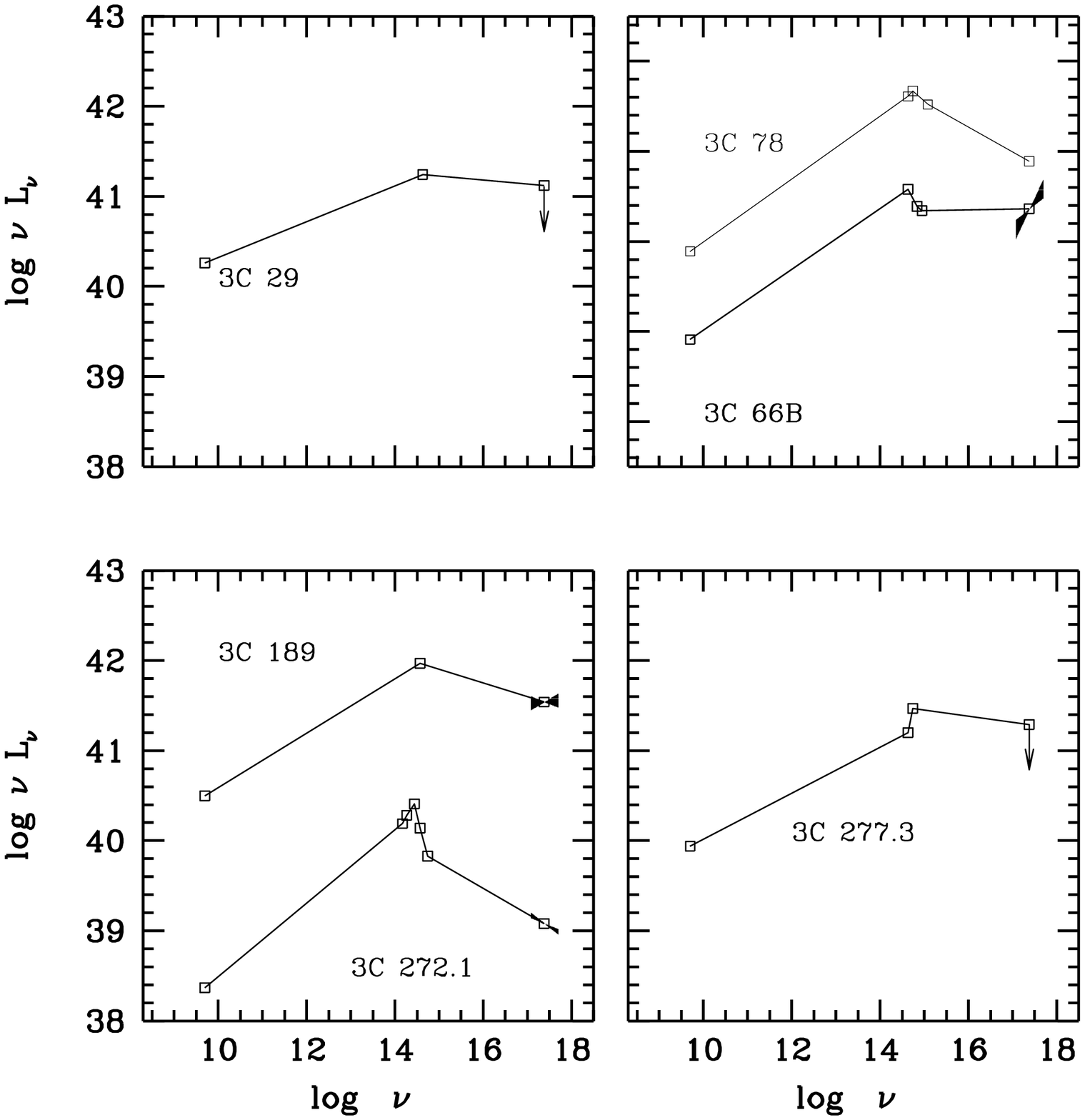},height=9.0truecm,angle=0}}
\centerline{\psfig{figure={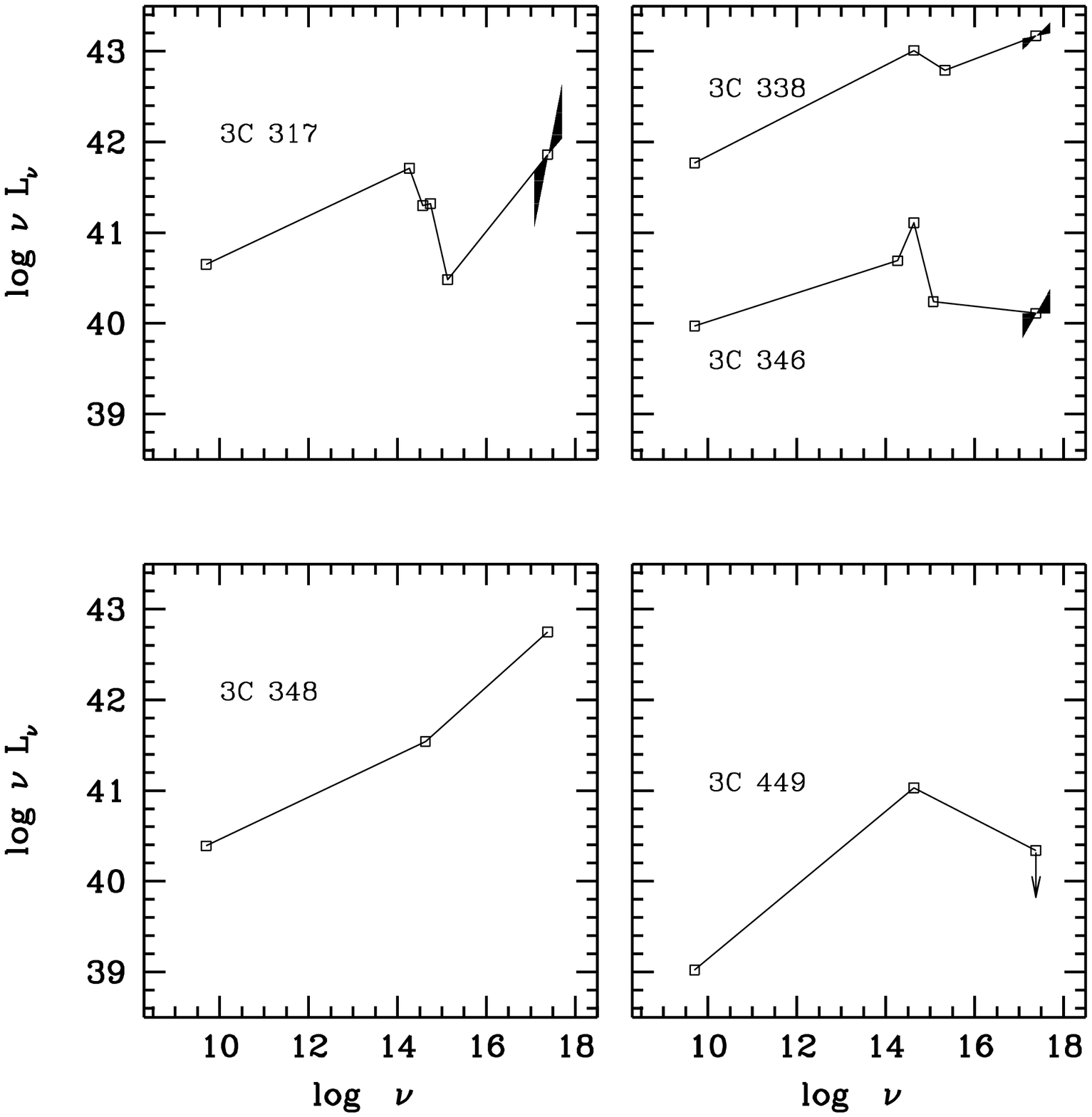},height=9.0truecm,angle=0}}
\caption{Spectral Energy Distribution of radio galaxies.
The values of 
$\nu L_{\nu}$ are in erg s$^{-1}$ }
\label{sed}
\end{figure*}

%Figure 2
\begin{figure*}[t]
\centerline{\psfig{figure=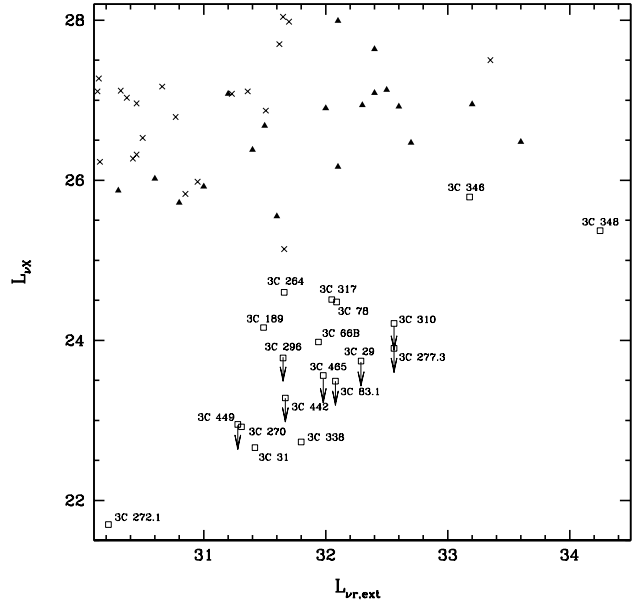,height=9.0truecm,angle=0}}
\caption{Core luminosities $L_{\nu {\rm X}}$ (erg s$^{-1}$ Hz$^{-1}$ at 1
keV) in the X-ray band of BL Lacs and radio galaxies (the latter ones
indicated by empty squares) vs the extended radio luminosity $L_{\nu {\rm r},
{\rm ext}}$
(erg s$^{-1}$ Hz$^{-1}$ at 1.4 GHz.). Filled triangles: LBL; crosses:
HBL}
\label{xext}
\end{figure*}

%Figure 3
\begin{figure*}[t]
\centerline{\psfig{figure=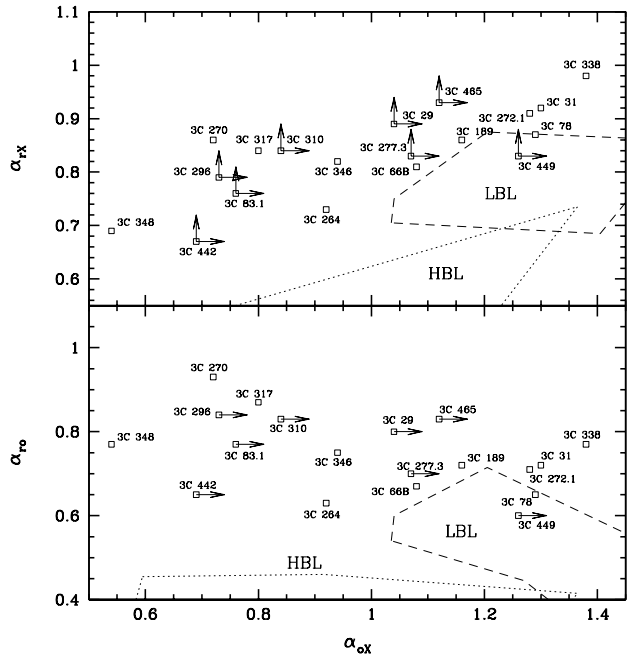,height=9.0truecm,angle=0}}
\caption{Plot of the broad-band spectral indices $\alpha_{\rm ro}$ and
$\alpha_{\rm rX}$ vs $\alpha_{\rm oX}$ for our sample, including also the  
sources of paper I (3C 288 is not included as it is found 
$\alpha_{\rm oX} > 0.34$). The regions occupied by LBL and HBL are 
from Fossati et al. (1998)}
\label{alpha}
\end{figure*}

%Figure 4
\begin{figure*}[t]
\centerline{\psfig{figure=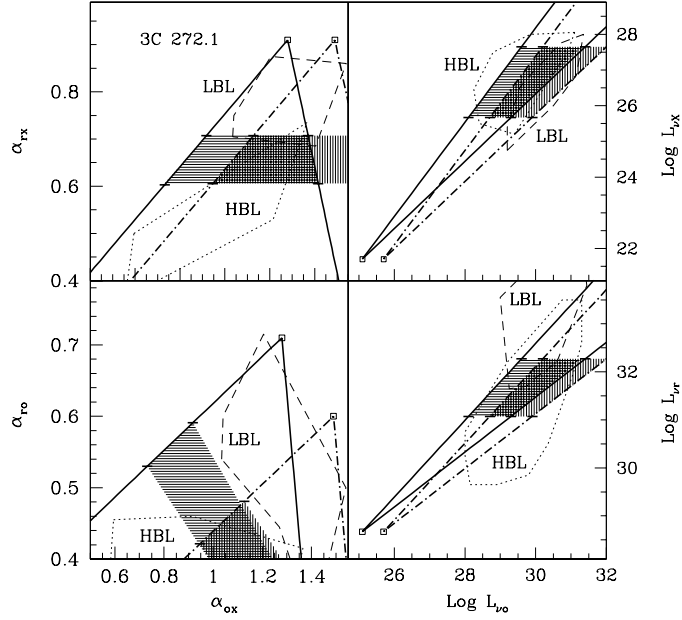,height=9.0truecm,angle=0}}
\caption{
Left panels: Beaming tracks of $\alpha_{\rm ro}$ and
$\alpha_{\rm rX}$ vs $\alpha_{\rm oX}$ for 3C 272.1 for no absorption (solid
lines) and 1.5 magnitudes of extinction (dash-dotted lines) at optical
wavelengths.  It has been assumed for the spectral indices
$\alpha_{\rm X=1.3}$, $\alpha_{\rm r}=0$, and $\alpha_{\rm o}= 0.5$ (leftwards tracks)
and $\alpha_{\rm o}= 1.5$ (rightwards tracks).  The shadowed regions limit
the values of allowed $R_{\delta}$ deduced from the comparison of
$L_{\nu {\rm X}}$ with $L_{\nu {\rm r},{\rm  ext}}$ (see Sect. 4).  
Right panels: Tracks
of $L_{\nu {\rm r}}$ and $L_{\nu {\rm X}}$ vs $L_{\nu {\rm o}}$ (erg s$^{-1}$ Hz$^{-1}$)
for the same parameters (upper tracks: $\alpha_{\rm o}= 0.5$, lower tracks:
$\alpha_{\rm o}= 1.5$)
}
\label{tr_272}
\end{figure*}

%Figure 5
\begin{figure*}[t]
\centerline{\psfig{figure=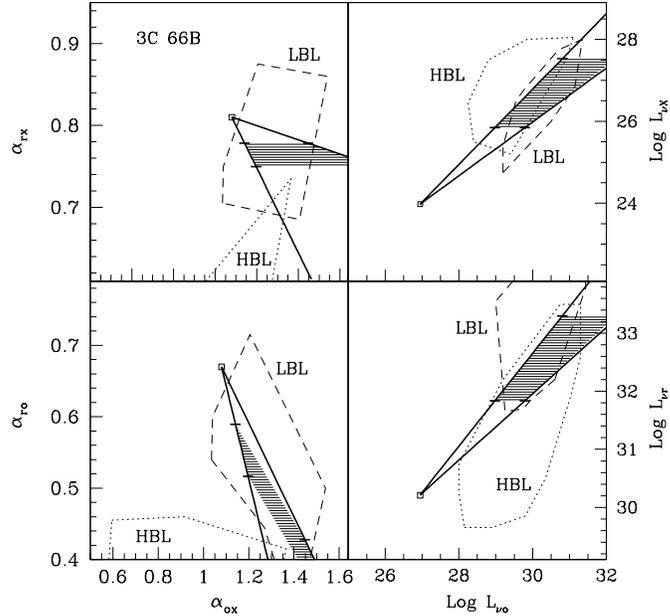,height=9.0truecm,angle=0}}
\caption{The same as Fig. 4 for 3C 66B, assuming $\alpha_{\rm X}=0.3$}
\label{tr_66B}
\end{figure*}

%Figure 6
\begin{figure*}[t]
\centerline{\psfig{figure=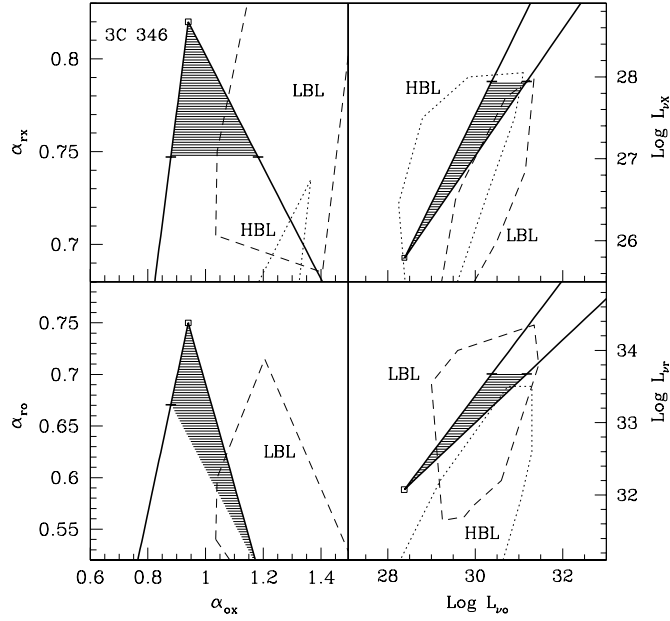,height=9.0truecm,angle=0}}
\caption{The same as Fig. 4 for 3C 346, assuming $\alpha_{\rm X}=0.7$}
\label{tr_346}
\end{figure*}

%Figure 7
\begin{figure*}[t]
\centerline{\psfig{figure=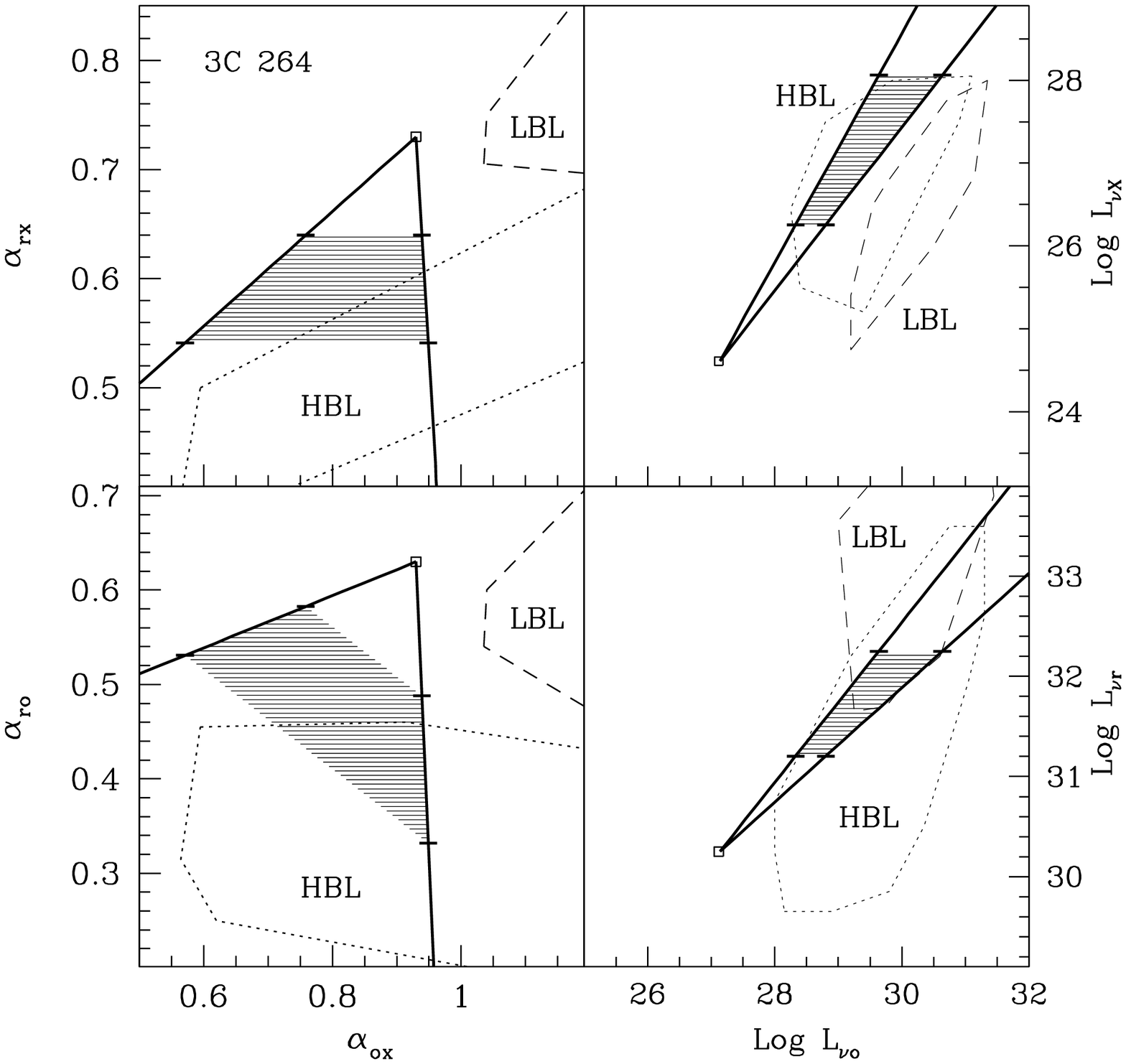,height=9.0truecm,angle=0}}
\caption{The same as Fig. 4 for 3C 264, assuming $\alpha_{\rm X}=1.45$}
\label{tr_264}
\end{figure*}

\appendix

\section{Notes on the X-ray properties of some sources}\label{A}
We summarize here the observations and the main results of 11 radio
galaxies observed with Rosat/PSPC, Chandra, BeppoSAX and ASCA. The
properties of the other six sources of the sample are discussed in
detail in HW99.

For the analysis of the Rosat/PSPC observations of 3C 29, 3C 66B, 3C
277.3, 3C 189 and 3C 317 (not yet analyzed elsewhere) we have followed
the same procedure discussed in paper I, i.e. we performed a spectral
analysis using composite models allowing for the presence of both a
thermal and a non-thermal (power-law) component.  To establish the
statistical significance of the non-thermal component, we have then
checked through an F-test when its inclusion improves the fit.  The
size of the region of photon extraction has been chosen to avoid as
much as possible contamination from the extended diffuse zone, and
taking into account the intensity of the sources and their position in
the detector FOV. In addition for targets offset more than
$20^{\prime}$ from the center of the FOV we have excluded the first 50
channels of energy, where the background flux largely increases. The
data have been processed using the Exsas package (version 1998) and
the reported errors are given at 1 $\sigma$ level for a single
parameter ($\Delta \chi^2=1$).  The uncertainties on the luminosities
(in the range 0.1 - 2.4 keV) are at the same level, but for two
parameters ($\Delta \chi^2=2.3$). The upper limits are given
at $3 \sigma$ levels, while no error is reported for sources where
thermal contamination is probable (for the details concerning the
statistical criteria for the other sources listed in Table 2, see the
quoted literature).

\noindent
{\bf 3C 29}. X-ray emission from this radio galaxy (observed in 1992
with an exposure of 15197 s), offset by $\approx 30^{\prime}$ in the
FOV (centered on the cluster A 119), has been detected with a $
\approx 3 \sigma$ confidence level (c. r.  $= 5 \pm 2 \times 10^{-3}$
s$^{-1}$). Assuming a power law spectrum with $\alpha_{\rm X} = 1.0$ we deduce 
an unabsorbed flux of $ 1.0 \times 10^{-13}$ erg
cm$^{-2}$ s$^{-1}$. The X-ray emission is consistent with that
expected from the correlation between the radio and X-ray core
luminosities (Canosa et al. 1999). To account for a possible
contribution from a thermal hot corona then, as for the data of HW99,
this emission is assumed as an upper limit to the core luminosity.

\noindent
{\bf 3C 31}. In paper I we derived, from Rosat/PSPC data, a nuclear flux
of $\approx 2.0 \times 10^{-13}$ erg cm$^{-2}$ s$^{-1}$, uncertain within 
a factor $\approx 2$ ($1 \sigma$) and without any estimate of the spectral 
index. From the Chandra observation of November 2000 (Hardcastle et al. 2002)
the inner region has been resolved in the pointlike nucleus and a jet, with 
comparable luminosities in the Rosat X-ray band. The spectrum is quite flat 
($\alpha_{\rm X} = 0.5^{+0.3}_{-0.2}$), with no evidence of a local absorption 
($N_{\rm H} = 1.5^{+2.2}_{-1.4} \times 10^{21}$ cm$^{-2}$). The shape of the SED 
is very similar to that of 3C 189 (see Fig. 1). The flux is $\approx 5 \times
 10^{-14}$  erg cm$^{-2}$ s$^{-1}$, i.e. $\approx 4$ times less than 
detected in Rosat but still consistent considering the statistical 
fluctuations and the jet contribution.

\noindent
{\bf 3C 66B}.  A Chandra observation of November 2000 has shown that
non-thermal flux originates from a point-like nucleus and a jet, which
are embedded in a diffuse, thermally emitting plasma. From the
spectral analysis,  an index
$1.03^{+0.07}_{-0.05} $ and a flux $2.5 \times 10^{-13}$ erg cm$^{-2}$
s$^{-1}$ (Hardcastle et al. 2001) have been obtained for the core emission.  
This radio galaxy falls in the FOV
of two pointed PSPC observations of August 1991 (24654 s) and August
1993 (31497 s): in both cases 3C 66B is offset by $38^{\prime}$ from
the center, and at $\approx 6^{\prime}$ from the blazar 3C 66A. To
avoid as much as possible the contamination from this nearby powerful
source, the target photons have been extracted from a ring with radius
of only $2.5^{\prime}$, and those of the background from a coronal
ring centered on 3C 66A with inner and outer radii of $\approx
3^{\prime}$ and $\approx 8^{\prime}$, respectively.  This implies an
underestimate of the total luminosity.  
To search for the
core emission we have performed spectral fits with a three components
model, thermal (from the diffuse region) plus two power law spectra
(from the jet and the core). Keeping fixed the parameters for the
thermal and jet emission as deduced from the Chandra data, and
assuming $N_{\rm H} = N_{\rm H,gal}$ ($=9.15 \times 10^{20}$ cm$^{-2}$) we
obtain for the two epochs : $\alpha_{\rm X} = 0.20^{+0.20}_{-0.25}$ and flux
$9.5 \times 10^{-13}$ erg cm$^{-2}$ s$^{-1}$ (1991), and $\alpha_{\rm X} =
0.30^{+0.30}_{-0.40}$ and flux $6.0 \times 10^{-13}$ erg cm$^{-2}$
s$^{-1}$ (1993). Assuming, for consistency with the Chandra results,
the largest allowed values of the spectral indices at $2 \sigma$
level of confidence (two related parameters, $\Delta \chi^2 = 4.6$),
$\alpha_{\rm X} = 0.75$ (1991) and $\alpha_{\rm X} = 0.85$ (1993), the fluxes are
$1.2 \times 10^{-12}$ erg cm$^{-2}$ s$^{-1}$ and $8.1 \times 10^{-13}$
erg cm$^{-2}$ s$^{-1}$ for the former and latter observation,
respectively.  This comparison of the Chandra and Rosat data indicates
that the core X-ray flux of 3C 66B has decreased by a factor $\sim 4$
in 10 years, with a remarkable softening of the spectrum. For the
multi-wavelength analysis we used the 1993 PSPC data, closer in time
to the HST observations.

\noindent
{\bf 3C 78}. The spectral analysis of a BeppoSAX observation of this
source (January 1997) supports the presence of non-thermal emission,
but without the estimate of the spectral index (Trussoni et
al. 1999). If we assume $\alpha_{\rm X} = 1.0$ we get a flux of $\approx 1.5
\times 10^{-13}$ erg cm$^{-2}$ s$^{-1}$, corresponding to a value of
the luminosity expected from the radio core emission.

\noindent
{\bf 3C 189}. From a Chandra observation of April 2000 (Worral et
al. 2001) the nuclear emission is fitted with a non-thermal spectrum
with $\alpha_{\rm X} = 1.1^{+0.1}_{-0.4}$, with flux $3.6 \times 10^{-13}$
erg cm$^{-2}$ s$^{-1}$.  Approximately the same energy flux is emitted
by a jet and from a diffuse thermal halo surrounding the central
region within a radius of $10^{\prime \prime}$.  To verify for a
possible variability we have analyzed the data of a Rosat/PSPC
observation of 1991. From a spectral fit, assuming a thermal + 
non-thermal components, we get a slightly softer spectrum, 
$\alpha_{\rm X} = 1.4^{+0.2}_{-0.2}$, but still
consistent with the Chandra observation. The resulting flux to be $
= 6.2 \times 10^{-13}$ erg cm$^{-2}$ s$^{-1}$; fixing $\alpha_{\rm X}=1.2$
the allowed range of the flux is $
3.0 - 6.7 \times 10^{-13}$ erg cm$^{-2}$ s$^{-1}$ at $2 \sigma$ level.
Then, if X-ray variability occurred in the nucleus of 3C 189 in the
period 1991 - 2000, its amplitude had to be less than a factor $\sim
1.8$. In the present analysis we have adopted the Chandra results.

\noindent
{\bf 3C 272.1}. The X-ray emission from the nucleus of M 84 detected
by Chandra (May 2000) has been fit with a power law spectrum with
$\alpha_{\rm X} = 1.3^{+0.1}_{-0.1}$, $N_{\rm H} = 2.7^{+0.3}_{-0.3} \times
10^{21}$ cm$^{-2}$ and an unabsorbed flux of $ 1.9 \times 10^{-13}$
erg cm$^{-2}$ s$^{-1}$ (Finoguenov \& Jones 2001). We remark that the
hydrogen column density implies an extinction of $\approx 1.4$
magnitudes at optical wavelengths, consistent with the HST data
(CCC99). In the Rosat energy band the core luminosity is $5.0 \times
10^{39}$ erg s$^{-1}$, in agreement with the data of HW99.

\noindent
{\bf 3C 277.3}. No X-ray emission has been detected at the position of this 
radio galaxy associated with  the  cluster Coma A. 
The pointing center was offset by $\approx
45^{\prime}$ in the FOV (pointed in 1991, c.r. $\simless \, 10^{-3}$ 
s$^{-1}$, exposure of 20345 s). Assuming a power law model 
with $\alpha_{\rm X} = 1$ this corresponds to 
an upper limit flux of $ \simless \, 4.2 \times 10^{-14}$  
erg cm$^{-2}$ s$^{-1}$, consistent with the radio/X-ray correlation of 
Canosa et al. (1999).

\noindent
{\bf 3C 317}. This radio galaxy is the main member of the cooling
flow cluster A 2052: its X-ray properties (from HRI 
observations) were  analyzed by Rizza et al. (2000) who did not find strong
evidence of a compact core. The PSPC flux from the inner 
region (c.r. $= 0.23 \pm 0.01$  s$^{-1}$, total exposure of 9240 s
from two pointings of 1992 and 1993) is consistent 
with the emission from a plasma with $T = 1.4^{+0.2}_{-0.1}$ keV and 
metallicity $= 0.60^{+0.15}_{-0.10}$ ($\chi^2_{\rm r} =1.2$), while a simple 
power law is not acceptable ($\chi^2_{\rm r} = 3$). If we fix the metallicity  
to 0.5 a better fit is obtained at 94 \% of confidence including a non-thermal 
component, that appears very flat ($\alpha_{\rm X} \approx 0.1$) even though 
the determination of the slope is very uncertain.
The flux results to be  $\approx 7.4 \times 10^{-13}$ erg cm$^{-2}$ 
s$^{-1}$. This object has been observed in September 2000 with the ACIS-S
detector on Chandra, for a total exposure of 36754 s. The core emission 
has been fit to an unabsorbed power law with  $\alpha_{\rm X} = 
1.00^{+0.15}_{-0.15}$ and flux $\approx 2.1 \times 10^{-13}$ erg cm$^{-2}$ 
s$^{-1}$ in the Rosat energy range (Blanton et al. 2002; no jet-like structure
has been detected). This emission  is $\sim 3.5$ times weaker than deduced in the 
1993/93 observations and has a steeper spectrum. However the spectral 
parameters obtained from the PSPC data are affected by large statistical 
fluctuations and we have  verified that they are consistent with the Chandra 
results at the $\approx 2 \sigma$
level. Then it is not possible to argue about a variability of the 
nuclear X-ray emission.  For the analysis of the SED we have adopted 
the 1992/1993 observations, closer in time to the HST data. 

\noindent
{\bf 3C 338}. A Chandra observation of the cluster A 2199 (December
1999) has detected a point-like source coinciding with the center of
the galaxy NGC 6166 (Di Matteo et al. 2001). The low count rate does
not provide a well constrained estimate of the parameters: for a power
law model we deduce $\alpha_{\rm X} = 0.54^{+0.56}_{-0.54}$ and flux $1.8
\times 10^{-14}$ erg cm$^{-2}$ s$^{-1}$ (in the Rosat energy band).
However it has been found that the value of the spectral slope rapidly
increases with the hydrogen column: for $N_{\rm H,loc} \simless \, 2 \times
10^{21}$ cm$^{-2}$), Di Matteo et al. obtain $\alpha_{\rm X} \sim 1.9$ and a
flux $\approx 2$ times larger.

\noindent
{\bf 3C 346}. The observations of Rosat/PSPC (1993) and ASCA (1995) of
this radio galaxy have been recently analyzed by Worrall \& Birkinshaw
(2001).  From the Rosat data it turns out that most of the emission
is point-like, unabsorbed (local $N_{\rm H} \simless \, 2 \times 10^{21}$
cm$^{-2}$) with $\alpha_{\rm X} = 0.69^{+0.16}_{-0.14}$ and flux $= 1.1
\times 10^{-12}$ erg cm$^{-2}$ s$^{-1}$, while from ASCA (two years
later) a similar spectral slope ($\alpha_{\rm X} = 0.73^{+0.17}_{-0.23}$) is
found, but a significant lower flux ($7.4 \times 10^{-13}$ erg
cm$^{-2}$ s$^{-1}$). For our analysis we have used the 1995 ASCA data.

\noindent
{\bf 3C 348}. From ASCA (1998) and Rosat data (PSPC in 1993, HRI in
1996) Siebert et al. (1999) detected non-thermal emission from the
core of this radio galaxy with very poor determination of the spectral
index ($\alpha_{\rm X} = 0.7^{+0.7}_{-0.3}$) and flux $\approx 3 \times
10^{-13}$ erg cm$^{-2}$ s$^{-1}$. This corresponds to $\sim 10\%$ of
the total extended thermal X-ray emission. These results have been
confirmed by an observation with BeppoSAX of March 1997 (Trussoni et
al. 2001).  It is worth noticing that 3C 348 is extremely bright at
high energies: the optical core luminosity is consistent with its
radio emission (CCC99), while the X-ray luminosity is $\approx$ ten
times higher than expected from the radio/X-ray correlation (Canosa et
al. 1999).

\end{document}